\documentclass[11pt]{article}
\usepackage{amsfonts,amsmath}
\def\bbt{\bibitem}
\def\be{\begin{equation}}
\def\en{\end{equation}}
\def\ber{\begin{eqnarray}}
\def\enr{\end{eqnarray}}
\def\nmb{ \nonumber\\}
\def\d{\partial}

\def\rbrc{\rbrace}
\def\lbrc{\lbrace}
\def\ov{\over }
\def\tld{\tilde}

\def\sgm{\sigma}
\def\Sgm{\Sigma}
\def\al{\alpha}
\def\bet{\beta}
\def\gm{\gamma}

\def\im{\imath}

\def\et{\eta}

\def\ups{\upsilon}

\def\dlt{\delta}

\def\bh{{\bf h}}

\def\bt{{\bf t}}

\def\be{{\bf e}}

\def\bmu{{\boldsymbol \mu}}

\begin{document}
%\nopagenumbers
%\rightline{Landau Tmp.}
\vskip 2 true cm

\centerline{\bf Free-field Representations and Geometry of some Gepner models.}

\vskip 1.5 true cm
\centerline{\bf S. E. Parkhomenko}
\centerline{Landau Institute for Theoretical Physics}
\centerline{142432 Chernogolovka, Russia}
\vskip 0.5 true cm
\centerline{spark@itp.ac.ru}
\vskip 1 true cm
\centerline{\bf Abstract}
\vskip 0.5 true cm

 The geometry of $k^{K}$ Gepner model, where $k+2=2K$ is 
investigated by free-field representation known as "$bc\bet\gm$"-system. 
Using this representation it is shown directly that internal sector of the model is given by 
Landau-Ginzburg $\mathbb{C}^{K}/\mathbb{Z}_{2K}$-orbifold. Then we consider the
deformation of the orbifold by marginal anti-chiral-chiral operator. Analyzing the holomorphic
sector of the deformed space of states we show that it has chiral de Rham complex
structure of some toric manifold, where toric dates are given by certain fermionic
screening currents. It allows to relate the Gepner model deformed by the marginal
operator to the $\sigma$-model on CY manifold realized as double cover of 
$\mathbb{P}^{K-1}$ with ramification along certain submanifold.

\vskip 10pt

"{\it PACS: 11.25Hf; 11.25 Pm.}"


\smallskip
\vskip 10pt
\centerline{\bf 1. Introduction}
\vskip 10pt

 Geometric aspects underlying
purely algebraic, Conformal Field Theory (CFT) construction of Gepner \cite{Gep} of the 
superstring vacua is an important and interesting area of study. It has two decades
history of research with a number of bright results. In consequence of this the relationship 
between the $\sgm$-models on Calabi-Yau (CY) manifolds and Gepner models has been clarified
essentially. For the review and the references on the original papers see \cite{Gre}.

 However the question how to relate directly the $\sgm$-model geometry to the algebraic
dates of Gepner's construction and when it is possible is still open.

 In the important work of Borisov \cite{B} the vertex operator algebra endowed with $N=2$
Virasoro superalgebra action has been constructed for each pair of dual reflexive polytopes
defining toric CY manifold. Thus, Borisov constructed directly holomorphic sector of the CFT 
from toric dates of CY manifold.
His approach is based essentially on the important 
work of Malikov, Schechtman and Vaintrob \cite{MSV} where a certain sheaf of vertex algebras 
which is called chiral de Rham complex has been introduced. Roughly speaking the construction 
of ~\cite{MSV} is a kind of free-field representation known as $"bc\bet\gm"$-system which is in case of 
Gepner model closely related with the Feigin and Semikhatov free-field representation ~\cite{FeS} of $N=2$ supersymmetric minimal models.  This circumstance is probably the key for understanding string geometry of Gepner models and their relationship to the $\sgm$-models on toric CY manifolds.

 The significant step in this direction has been made in the paper~\cite{GorbM} where
the vertex algebra of certain Landau-Ginzburg (LG) orbifold has been related to chiral de Rham
complex of toric CY manifold by a spectral sequence. The CY manifold has been realized
as an algebraic surface degree $K$ in the projective space $\mathbb{P}^{K-1}$ and 
one of the key points of ~\cite{GorbM} is that the free-field representation of the corresponding LG orbifold is given by $K$ copies of $N=2$ minimal model free-field representation of \cite{FeS}.

 The Gepner model can be characterized by $K$-dimensional vector 
\ber
\bmu=(\mu_{1},...,\mu_{K})
\label{1.vectmu}
\enr
where 
\ber
\mu_{i}=2,3,..., \ i=1,...,K
\label{1.mu}
\enr
define the central charges of the individual $N=2$ minimal models
\ber
c_{i}=3(1-{2\ov\mu_{i}})
\label{1.c}
\enr 
In what follows the $\mu_{i}$ will be specified by
\ber
\bmu=(\mu,\mu,...,\mu)
\label{1.vectmu1}
\enr
so the total central charge of the model is 
\ber
c=\sum_{i=1}^{K}c_{i}=3K(1-{2\ov\mu})
\label{1.ctot}
\enr
There are two cases when the central charge is integer and multiple of 3
\ber
\mu=K, \ 2K
\label{1.CYcase}
\enr
The geometry underlying the first case has been investigated in \cite{GorbM}.

 In the second case the geometry is more interesting. The total central charge is
\ber
c=3(K-1)
\label{1.secndc}
\enr
and hence the complex dimension of the compact manifold is $K-1$.
I am going to show in this note that the internal geometry of Gepner model
corresponds in this case to the $\sigma$-model on the CY manifold which double covers the 
$\mathbb{P}^{K-1}$ with ramification along certain submanifold. It means in particular
that center of mass of the string is allowed to move only along the base
$\mathbb{P}^{K-1}$ but with some twisted sectors added along the fiber
of the double cover.

 One can generalize the second case and consider the models where 
\ber
\mu=3K,4K,...
\label{1.MK}
\enr
Though the total central charge is no longer integer and these models can not be
used as the models of superstring compactification, the orbifold projection consistent
with modular invariance still exits \cite{KYY} which makes them to be interesting $N=2$ supersymmetic
models of CFT from the geometric point of view. The geometry of these models has been
investigated partly in \cite{P1}.

 In Section 2 I represent a collection of known facts on the $N=2$ minimal models,
fix the notations and briefly remind the Gepner's construction of the partition
function in the internal sector of the Gepner model. In Section 3 the free-field representation
of \cite{FeS} is used to relate the model with LG $\mathbb{C}^{K}/\mathbb{Z}_{2K}$-orbifold.

In Sect.4 the resolution of the orbifold singularity in chiral sector is considered. It is given by
adding some new fermionic screening charge coming from the twisted sector of Gepner model. 
We show that this additional screening charge together 
with the old charges define the toric dates of $O(K)$-bundle total space over the $\mathbb{P}^{K-1}$
as well as the potential on this space. The chiral sector space of states of the model has the
chiral de Rham complex structure on the $O(K)$-bundle total space restricted to zeroes of the gradient of the potential. Then we consider the rest of the orbifold group action on the space of 
states and relate the model with $\sgm$-model on a CY manifold which double covers the projective space $\mathbb{P}^{K-1}$. 
%At the end we breafly discuss the possible interpretation of the Gepner
%models as $\sgm$-models on $\mathbb{P}^{K-1}$ with fluxes.

\vskip 10pt
\centerline{\bf 2. The internal sector partition function of the Gepner models.}
\vskip 10pt

 In this section we remind the construction of the partition function of the Gepner model
in the internal sector. To be more specific the Ramound-Ramound (RR) partition function of the 
internal sector will be important for the geometry investigation. But as a preliminary
we represent a collection of known facts on the $N=2$ minimal models and fix the notations.

\leftline {\bf 2.1. The products of $N=2$ minimal models.}

 The tensor product of $K$ $N=2$ unitary minimal models can be characterized
by $K$-dimensional vector $\bmu=(\mu_{1},...,\mu_{K})$,
where $\mu_{i}\geq 2$ being integer defines the central charge of the individual model
by $c_{i}=3(1-{2\ov\mu_{i}})$. 
For each individual minimal model we denote by $M_{h,t}$ the irreducible unitary 
$N=2$ Virasoro superalgebra representation in NS sector and denote by 
$\chi_{h,t}(q,u)$ the character of the representation
\ber
\chi_{h,t}(q,u)=Tr_{h,t}(q^{L[0]-{c\ov 24}}u^{J[0]})
\label{2.chi}
\enr
where
$h=0,...,\mu-2$ and $t=0,...,h$. 
There are the following important automorphisms of the irreducible modules and characters
~\cite{FeS},~\cite{FeSST}.
\ber
M_{h,t}\equiv M_{\mu-h-2,t-h-1}, \
\chi_{h,t}(q,u)=\chi_{\mu-h-2,t-h-1}(q,u),
\label{2.reflect}
\enr
\ber
M_{h,t}\equiv M_{h,t+\mu}, \ \chi_{h,t+\mu}(q,u)=
\chi_{h,t}(q,u),
\label{2.oddper}
\enr
where $\mu$ is odd and
\ber
M_{h,t}\equiv M_{h,t+\mu}, \ \chi_{h,t+\mu}(q,u)=
\chi_{h,t}(q,u),\ h\neq {\mu\ov2}-1, \nmb
M_{h,t}\equiv M_{h,t+{\mu\ov2}}, \
\chi_{h,t+{\mu\ov 2}}(q,u)=\chi_{h,t}(q,u),\ h={\mu\ov2}-1,
\label{2.evper}
\enr
where $\mu$ is even.
In what follows we extend the set of admissible $t$: 
\ber
t=0,...,\mu-1
\label{2.textend}
\enr
using the automorphisms above.

 The parameter $t\in \mathbb{Z}$ labels the spectral flow automorphisms \cite{SS} of $N=2$
Virasoro superalgebra in NS sector
\ber
G^{\pm}[r]\rightarrow G_{t}^{\pm}[r]\equiv U^{t}G^{\pm}[r]U^{-t}\equiv G^{\pm}[r\pm t], \nmb
L[n]\rightarrow L_{t}[n]\equiv U^{t}L[n]U^{-t}\equiv L[n]+t J[n]+t^{2}{c\ov 6}\dlt_{n,0},
\nmb
J[n]\rightarrow J_{t}[n]\equiv U^{t}J[n]U^{-t}\equiv J[n]+t {c\ov 3}\dlt_{n,0},
\label{2.flow}
\enr
where $U^{t}$ denotes the spectral flow operator generating twisted sectors. Here $r$ is half-integer for the modes of the spin-$3/2$ fermionic currents $G^{\pm}(z)$ while $n$ is integer for the modes of stress-energy tensor $T(z)$ and $U(1)$-current $J(z)$ of the $N=2$ Virasoro superalgebra. 
So allowing $t$ to be half-integer we recover the irreducible
representations and characters in the $R$ sector.

 The $N=2$ Virasoro superalgebra generators in the product of minimal models are given by the sums of
generators of each minimal model
\ber
G^{\pm}[r]=\sum_{i}G^{\pm}_{i}[r], \nmb
J[n]=\sum_{i}J_{i}[n], \
T[n]=\sum_{i}T_{i}[n],
\nmb
c=\sum_{i}3(1-{2\ov\mu_{i}})
\label{2.Vird}
\enr
This algebra is obviously acting in the tensor product
$M_{\bh,\bt}=\otimes_{i=1}^{K}M_{h_{i},t_{i}}$ of the irreducible $N=2$ Virasoro superalgebra 
representations of each individual model. We use the similar notation for the corresponding 
product of characters
\ber
\chi_{\bh,\bt}(q,u)=\prod_{i=1}^{K}\chi_{h_{i},t_{i}}(q,u)
\label{2.prodchi}
\enr

\leftline {\bf 2.2. The partition function of the internal sector.}

 In what follows the characters with fermionic number operator insertion will be important
\ber
\tld{\chi}_{h_{i},t_{i}}(q,u)=Tr_{h_{i},t_{i}}((-1)^{F}q^{L_{i}[0]-{c_{i}\ov 24}}u^{J_{i}[0]}).
\label{2.tldchi}
\enr

 The internal sector partition function of the Gepner model in RR-sector is given by
\ber
Z(q,\bar{q},u,\bar{u})=
{1\ov 2K2^{K}}\sum_{n,m}^{2K-1}\prod_{i=1}^{K}
\sum_{h_{i},t_{i}}\tld{\chi}_{h_{i},t_{i}+n+{1\ov 2}}(\tau,\ups +m)
\tld{\chi}^{*}_{h_{i},+t_{i}+{1\ov 2}}(\tau,\ups )
\label{2.ZRR}
\enr
where $q=\exp{[\im 2\pi\tau]}$, $u=\exp{[\im 2\pi\ups]}$ and $*$ denotes the complex conjugation.
The summation over $n$ is due to the spectral flow twisted sector
generated by the product of spectral flow operators $\prod_{i=1}^{K}U_{i}^{n}$.
The summation over $m$ corresponds to the projection on the $\mathbb{Z}_{2K}$-invariant states
with respect to the operator $\exp{[\im 2\pi J[0]]}$.
Thus it is $\mathbb{Z}_{2K}$-orbifold partition function in RR-sector with periodic spin structure
along the both cycles of the torus.

\vskip 10pt
\centerline {\bf 3. Free-field representations and LG orbifold geometry of Gepner models.}
\vskip 10pt

 In this section we relate the Gepner models to the LG orbifolds 
$\mathbb{C}^{K}/\mathbb{Z}_{2K}$ using essentially the free-field construction of 
irreducible representations of $N=2$ minimal models
found by Feigin and Semikhatov in \cite{FeS}. 

\leftline{\bf 3.1. Free-field realization of $N=2$ minimal model.}

 Let $X(z), X^{*}(z)$ be the free bosonic fields and $\psi(z), \psi^{*}(z)$ be the
free fermionic fields (in the left-moving sector) 
so that its OPE's are given by
\ber
X^{*}(z_{1})X(z_{2})=\ln(z_{12})+reg.,\nmb
\psi^{*}(z_{1})\psi(z_{2})=z_{12}^{-1}+reg,
\label{3.ope}
\enr
where $z_{12}=z_{1}-z_{2}$. For an arbitrary number
$\mu$ the currents of $N=2$ super-Virasoro
algebra are given by
\ber
G^{+}(z)=\psi^{*}(z)\d X(z) -{1\ov \mu} \d \psi^{*}(z), \
G^{-}(z)=\psi(z) \d X^{*}(z)-\d \psi(z), \nmb
J(z)=\psi^{*}(z)\psi(z)+{1\ov \mu}\d X^{*}(z)-\d X(z), \nmb
T(z)=\d X(z)\d X^{*}(z)+
{1\ov 2}(\d \psi^{*}(z)\psi(z)-\psi^{*}(z)\d \psi(z))-\nmb
{1\ov 2}(\d^{2} X(z)+{1\ov \mu}\d^{2} X^{*}(z)),
\label{3.min}
\enr
and the central charge is
\ber
c=3(1-{2\ov \mu}).
\label{3.cent}
\enr

  As usual, the fermions are expanded into the
half-integer modes in NS sector and they are expanded into integer modes in R sector
\ber
\psi(z)=\sum_{r}\psi[r]z^{-{1\ov 2}-r},\
\psi^{*}(z)=\sum_{r}\psi^{*}[r]z^{-{1\ov 2}-r},\
G^{\pm}(z)=\sum_{r}G^{\pm}[r]z^{-{3\ov 2}-r},
\label{3.NSR}
\enr
The bosons are expanded in both sectors into the integer
modes:
\ber
\d X(z)=\sum_{n\in Z}X[n]z^{-1-n},\
\d X^{*}(z)=\sum_{n\in Z}X^{*}[n]z^{-1-n},\nmb
J(z)=\sum_{n\in Z}J[n]z^{-1-n},\
T(z)=\sum_{n\in Z}L[n]z^{-2-n}.
\label{3.NSRb}
\enr

 In NS sector $N=2$ Virasoro superalgebra is acting naturally in Fock module
$F_{p,p^{*}}$ generated by the fermionic
operators $\psi^{*}[r]$, $\psi[r]$, $r<{1\ov2}$, and bosonic operators
$X^{*}[n]$, $X[n]$, $n<0$ from the vacuum state $|p,p^{*}>$ such that
\ber
\psi[r]|p,p^{*}>=\psi^{*}[r]|p,p^{*}>=0, r\geq {1\ov 2},\nmb
X[n]|p,p^{*}>=X^{*}[n]|p,p^{*}>=0, n\geq 1, \nmb
X[0]|p,p^{*}>=p|p,p^{*}>, \
X^{*}[0]|p,p^{*}>=p^{*}|p,p^{*}>.
\label{3.vac}
\enr
It is a primary state with respect
to the $N=2$ Virasoro algebra
\ber
G^{\pm}[r]|p,p^{*}>=0, r>0, \nmb
J[n]|p,p^{*}>=L[n]|p,p^{*}>=0, n>0, \nmb
J[0]|p,p^{*}>={j\ov \mu}|p,p^{*}>=0, \nmb
L[0]|p,p^{*}>={h(h+2)-j^{2}\ov 4\mu}|p,p^{*}>=0,
\label{3.hwv}
\enr
where $j=p^{*}-\mu p$, $h=p^{*}+\mu p$.

 When $\mu-2$ is integer and non negative the Fock module is highly reducible representation 
of $N=2$ Virasoro algebra.

The irreducible module $M_{h,j}$ is given by cohomology of some complex building up from 
Fock modules. This complex has been constructed in ~\cite{FeS}. Let us consider first free-field 
construction for the chiral module $M_{h,0}$. In this case the complex
(which is known due to Feigin and Semikhatov as butterfly resolution)
can be represented by the following diagram
\ber
\begin{array}{ccccccccccc}
&&\vdots &\vdots &&&&&&\\
&&\uparrow &\uparrow &&&&&&\\
\ldots &\leftarrow &F_{1,h+\mu} &\leftarrow
F_{0,h+\mu}&&&&&&\\
&&\uparrow &\uparrow &&&&&&\\
\ldots &\leftarrow &F_{1,h} &\leftarrow F_{0,h}&&&&&&\\
&&&&\nwarrow&&&&&\\
&&&&&F_{-1,h-\mu}&\leftarrow &F_{-2,h-\mu}&\leftarrow&\ldots\\
&&&&&\uparrow &&\uparrow&\\
&&&&&F_{-1,h-2\mu}&\leftarrow &F_{-2,h-2\mu}&\leftarrow&\ldots\\
&&&&&\uparrow &&\uparrow &&\\
&&&&&\vdots &&\vdots &&
\end{array} \nmb
\label{3.but}
\enr
The horizontal arrows in this diagram are given by the action of
\ber
Q^{+}=\oint dz S^{+}(z), \ S^{+}(z)=\psi^{*}\exp(X^{*})(z),
\label{3.chrg+}
\enr
The vertical arrows are given by the action of 
\ber
Q^{-}=\oint dz S^{-}(z), \ S^{-}(z)=\psi\exp(\mu X)(z),
\label{3.chrg-}
\enr
The diagonal arrow at the middle of butterfly resolution
is given by the action of $Q^{+}Q^{-}$. It is a complex due to the following properties
screening charges $Q^{\pm}$
\ber
(Q^{+})^{2}=(Q^{-})^{2}=\{Q^{+},Q^{-}\}=0.
\label{3.BRST}
\enr

 The main statement of ~\cite{FeS} is that the complex (\ref{3.but}) is exact
except at the $F_{0,h}$ module, where the cohomology is given by
the chiral module $M_{h,0}$.

 To get the resolution for the irreducible module $M_{h,t}$
one can use the observation ~\cite{FeS} that all irreducible modules can be obtained
from the chiral module $M_{h,0}$, $h=0,...,\mu-2$ by the spectral flow
action $U^{-t}, t=1,...,\mu-1$. The spectral flow action on the free
fields can be easily described if we bosonize fermions $\psi^{*}, \psi$
\ber
\psi(z)=\exp(-\phi(z)), \ \psi^{*}(z)=\exp(\phi(z)).
\label{3.fbos}
\enr
and introduce spectral flow vertex operator
\ber
U^{t}(z)=\exp(-t(\phi+{1\ov \mu}X^{*}-X)(z)).
\label{3.vflow}
\enr

Using the resolution (\ref{3.but}) and the spectral flow we obtain the 
following expression for the character \cite{FeSST}
\ber
\chi_{h,-t}(u,q)=q^{{h\ov 2\mu}+{c\ov 6}t^{2}+{th\ov\mu}-{c\ov 24}}q^{{1-\mu\ov 8}}
u^{{h\ov\mu}+{ct\ov 3}}
({\et(q^{\mu})\ov\et(q)})^{3}
\nmb
\prod_{n=0}{(1+uq^{{1\ov 2}+t+n})\ov (1+u^{-1}q^{-{1\ov 2}-t+n\mu})}
{(1+u^{-1}q^{{1\ov 2}-t+n})\ov (1+uq^{{1\ov 2}+t+(n+1)\mu})}
{(1-q^{n+1})\ov (1-q^{(n+1)\mu})}
\nmb
\prod_{n=0}{(1-q^{-1-h+n\mu})\ov (1+uq^{-{1\ov 2}-h+t+n\mu})}
{(1-q^{1+h+(n+1)\mu})\ov (1+u^{-1}q^{{1\ov 2}+h-t+(n+1)\mu})}
\label{3.char}
\enr
where
\ber
\et(q)=q^{{1\ov 24}}\prod_{n=1}(1-q^{n})
\label{3.Ded}
\enr
 
 The resolutions and irreducible modules in R sector are
generated from the resolutions and modules in NS sector by the
spectral flow operator $U^{{1\ov2}}$.

\leftline {\bf 3.2.Free-field realization of the product of
minimal models.}

 It is clear how to generalize the free-field representation to the case of
tensor product of $K$ $N=2$ minimal models.
One has to introduce (in the left-moving sector)
the free bosonic fields $X_{i}(z), X^{*}_{i}(z)$ and free
fermionic fields $\psi_{i}(z), \psi^{*}_{i}(z)$, $i=1,...,K$
so that its singular OPE's are given by (\ref{3.ope}).
The $N=2$ superalgebra Virasoro currents for each of the models are given by (\ref{3.min}).
To describe the products of irreducible representations $M_{\bh,\bt}$ we introduce
the fermionic screening currents and their charges
\ber
S^{+}_{i}(z)=\psi^{*}_{i}\exp(X^{*}_{i})(z), \ 
S^{-}_{i}(z)=\psi_{i}\exp(\mu_{i}X_{i})(z), \
Q^{\pm}_{i}=\oint dz S^{\pm}_{i}(z).
\label{3.chrgi}
\enr
Then the module $M_{\bh,0}$ is given by the cohomology of the product
of butterfly resolutions (\ref{3.but}). The resolution
of the module $M_{\bh,\bt}$ is generated by the spectral flow operator
$U^{\bt}=\prod_{i}U_{i}^{t_{i}}$, $t_{i}=1,...,\mu_{i}-1$, where
$U_{i}^{t_{i}}$ is the spectral flow operator from the $i$-th minimal model (\ref{3.vflow}). 
Allowing $t_{i}$ to be half-integer we generate the corresponding objects in R sector.
In what follows we consider the case $\mu_{1}=...=\mu_{K}=2K$.

\leftline {\bf 3.3. LG orbifold geometry of Gepner models.}

 The holomorphic factor of the space of states of the model (\ref{2.ZRR}) in R-sector
is given also by cohomology of the complex. It is an orbifold of the complex which is the sum 
of butterfly resolutions for the modules $M_{\bh,\bt}$. 
The cohomology of this complex can be calculated by two steps.

 At first step we take the cohomology wrt the operator 
\ber
Q^{+}=\sum_{i=1}^{K}Q^{+}_{i}
\label{3.Qplus}
\enr
It is generated by $bc\bet\gm$ system of fields
\ber
a_{i}(z)=\exp{[X_{i}]}(z),\ \al_{i}(z)=\psi_{i}\exp{[X_{i}]}(z),
\nmb
a^{*}_{i}(z)=(\d X_{i}^{*}-\psi_{i}\psi^{*}_{i})
\exp{[-X_{i}]}(z), \
\al^{*}_{i}(z)=\psi^{*}_{i}\exp{[-X_{i}]}(z)
\label{3.btgm}
\enr
The singular operator product expansions of these fields are
\ber
a^{*}_{i}(z_{1})a_{j}(z_{2})=z_{12}^{-1}\dlt_{ij}+...,
\nmb
\al^{*}_{i}(z_{1})\al_{j}(z_{2})=z_{12}^{-1}\dlt_{ij}+....
\label{3.btgm1}
\enr
In terms of the fields (\ref{3.btgm}) the N=2 Virasoro superalgebra currents (\ref{2.Vird})
are given by
\ber
G^{-}=\sum_{i}\al_{i}a^{*}_{i}, \
G^{+}=\sum_{i}(1-{1\ov 2K})\al^{*}_{i}\d
a_{i}-{1\ov 2K}a_{i}\d\al^{*}_{i}, \nmb
J=\sum_{i}(1-{1\ov 2K})\al^{*}_{i}\al_{i}+{1\ov 2K}a_{i}a^{*}_{i},\nmb
T=\sum_{i}{1\ov 2}((1+{1\ov 2K})\d\al^{*}_{i}\al_{i}-
(1-{1\ov 2K})\al^{*}_{i}\d\al_{i})+
(1-{1\ov 4K})\d a_{i}a^{*}_{i}-
{1\ov 4K}a_{i}\d a^{*}_{i}
\label{3.btgmvir}
\enr
Notice that zero mode $G^{-}[0]$ is acting in the space of states generated by $bc\bet\gm$ system
of fields similar to the de Rham differential action in the de Rham complex of $\mathbb{C}^{K}$.
Due to this observation and taking into account (\ref{3.btgm1}) one can make the following
geometric interpretation of the fields (\ref{3.btgm}).
The fields $a_{i}(z)$ correspond to the coordinates $a_{i}$ on the complex space
$\mathbb{C}^{K}$, the fields $a^{*}_{i}(z)$ correspond to the operators ${\d \ov \d a_{i}}$.
The fields $\al_{i}(z)$ correspond to the differentials $da_{i}$, while $\al^{*}_{i}(z)$
correspond to the conjugated to $da_{i}$. 

 The next important property is the behaviour of the $bc\bet\gm$ system under the local change of
coordinates on $\mathbb{C}^{K}$ \cite{MSV}. For each new set of coordinates
\ber
b_{i}=g_{i}(a_{1},...,a_{K}), \
a_{i}=f_{i}(b_{1},...,b_{K})
\label{3.coordtr}
\enr
the isomorphic $bc\bet\gm$ system of fields is given by
\ber
b_{i}(z)=g_{i}(a_{1}(z),...,a_{K}(z)),
\nmb
\bet_{i}(z)={\d g_{i}\ov \d a_{j}}(a_{1}(z),...,a_{K}(z))\al_{j}(z),
\nmb
\bet^{*}_{i}(z)={\d f_{j}\ov \d b_{i}}(a_{1}(z),...,a_{K}(z))\al^{*}_{j}(z),
\nmb
b^{*}_{i}(z)={\d f_{j}\ov \d b_{i}}(a_{1}(z),...,a_{K}(z))a^{*}_{j}(z)+
\nmb
{\d^{2} f_{k}\ov\d b_{i}\d b_{j}}{\d g_{j}\ov\d a_{n}}(a_{1}(z),...,a_{K}(z))\al^{*}_{k}(z)\al_{n}(z)
\label{3.coordtr1}
\enr
where the normal ordering of the fields is implied.
It endows the $bc\bet\gm$ system (\ref{3.btgm})
with the structure of sheaf known as chiral de Rham complex due to \cite{MSV}. 

All these properties provide the geometric 
meaning to the algebraic construction of Gepner model. Indeed, it was shown by
Borisov in general toric setup \cite{B} that the screening charges $Q^{+}_{i}$ 
determine the toric dates of some toric manifold and the cohomology of the differentail
(\ref{3.Qplus}) gives the sections of chiral de Rham complex on this manifold.
In our case this manifold is $\mathbb{C}^{K}$ and chiral de Rham complex on this space
is generated by $bc\bet\gm$ system (\ref{3.btgm}).

 The charges of the fields (\ref{3.btgm}) are given by
\ber
J(z_{1})a_{i}(z_{2})=z_{12}^{-1}{1\ov 2K}a_{i}(z_{2})+r.,
\
J(z_{1})a^{*}_{i}(z_{2})=-z_{12}^{-1}{1\ov 2K}a^{*}_{i}(z_{2})+r.,
\nmb
J(z_{1})\al_{i}(z_{2})=-z_{12}^{-1}(1-{1\ov 2K})\al_{i}(z_{2})+r.,\
J(z_{1})\al^{*}_{i}(z_{2})=z_{12}^{-1}(1-{1\ov 2K})\al^{*}_{i}(z_{2})+r.
\label{3.jchrg}
\enr

 Hence, making the projection on $\mathbb{Z}_{2K}$-invariant states and adding twisted sectors 
generated by $\prod_{i=1}^{K}(U_{i})^{n}$ we obtain toric construction of the chiral 
de Rham complex of the orbifold $\mathbb{C}^{K}/\mathbb{Z}_{2K}$.
The chiral de Rham complex on the orbifold has recently been introduced in \cite{FrSc}.

 The second step in the cohomology calculation is given by
the cohomology with respect to the differential
$Q^{-}=\sum_{i=1}^{K}Q^{-}_{i}$.
This operator survives the orbifold projection and its expression in terms of fields
(\ref{3.btgm}) is
\ber
Q^{-}=\oint dz \sum_{i=1}^{K}\al_{i}(a_{i})^{2K-1}
\label{3.Qminus1}
\enr
Therefore the second step of cohomology calculation gives the restriction of the chiral de Rham
complex to the points $dW=0$ of the potential
\ber
W=\sum_{i=1}^{K}(a_{i})^{2K}
\label{3.LG}
\enr
Thus the total space of states is the space of states of LG orbifold 
$\mathbb{C}^{K}/\mathbb{Z}_{2K}$
whose partition function in RR-sector is given by (\ref{2.ZRR}). 

\vskip 10pt
\centerline {\bf 4. LG/sigma-model correspondence conjecture.}
\vskip 10pt

 In this section we relate the LG orbifold $\mathbb{C}^{K}/\mathbb{Z}_{2K}$
to the $\sgm$-model on CY manifold which double cover the space $\mathbb{P}^{K-1}$.
The relation appears when we deform LG-orbifold by marginal operator making the orbifold singularity
resolution. According to the construction  
\cite{B},~\cite{GorbM}, the orbifold singularity resolution in holomorphic sector
is given by supplementary screening charges.

\vskip 10pt
\leftline{\it 4.1. $K=2$ example.}

Let us consider first this procedure in the simplest example $K=2$.
In this case we add to the charges $Q^{+}_{1,2}$ the screening charge
\ber
D_{orb}=
\oint dz {1\ov 2}(\psi^{*}_{1}+\psi^{*}_{2})\exp({1\ov 2}(X^{*}_{1}+X^{*}_{2}))(z) 
\label{4.Dorb}
\enr
It is easy to check that this operator commutes with the total $N=2$ Virasoro superalgebra
currents (\ref{2.Vird}) and commutes also with the operators $Q^{-}_{i}$. The corresponding fermionic screening current is the holomorphic (chiral) factor of the anti-chiral-chiral marginal field \cite{LVW}, \cite{Gre}, coming from the twisted sector. The fermionic operators
\ber
D^{+}_{n}=\oint dz ({2-n\ov 4}\psi^{*}_{1}+{2+n\ov 4}\psi^{*}_{2})
\exp({2-n\ov 4}X^{*}_{1}+{2+n\ov 4}X^{*}_{2})(z),\ n=-1,1
\label{4.Qn}
\enr 
also commute with $N=2$ Virasoro algebra and $Q^{-}_{i}$ but they do not appear as marginal 
operators of the model because they should come from twisted sectors which are not exist 
in the model (see (\ref{2.ZRR})).

 Following the construction of Borisov we associate to the set of screening charges
$Q^{+}_{1},Q^{+}_{2},D_{orb}$ the fan \cite{Ful} consisting of two 2-dimensional cones $\sgm_{1}$ and $\sgm_{2}$, 
generated in the lattice
$({1\ov 2}\mathbb{Z})^{2}$ by the vectors $(e_{1},{1\ov 2}(e_{1}+e_{2}))$ and vectors
$(e_{2},{1\ov 2}(e_{1}+e_{2}))$ correspondingly. To each of the cones $\sgm_{i}$
the $bc\bet\gm$ system of fields is related by the cohomology of 
the differential $Q^{+}_{i}+D_{orb}$, $i=1,2$. This is the first step of cohomology calculation. 

 One can show that these two systems generate the space of sections of the chiral de Rham
complex on the open sets of the standard covering of the total space of $O(2)$-bundle 
over $\mathbb{P}^{1}$. 

 Indeed, one can split the first step of cohomology calculation into 2
substeps. At the first substep we take $Q^{+}_{1}+D_{orb}$-cohomology.
It is given by the following $bc\bet\gm$ fields
\ber
b_{0}(z)=\exp{[2X_{2}]}(z),
\
\bet_{0}(z)=2\psi_{2}\exp{[2X_{2}]}(z),
\nmb
b^{*}_{0}(z)=({1\ov 2}(\d X_{1}^{*}+\d X_{2}^{*})-
\psi_{2}(\psi^{*}_{1}+\psi^{*}_{2}))
\exp{[-2X_{2}]}(z), 
\
\bet^{*}_{0}(z)={1\ov 2}(\psi^{*}_{1}+\psi^{*}_{2})\exp{[-2X_{2}]}(z),
\nmb
b_{1}(z)=\exp{[X_{1}-X_{2}]}(z),
\
\bet_{1}(z)=(\psi_{1}-\psi_{2})\exp{[X_{1}-X_{2}]}(z),
\nmb
b^{*}_{1}(z)=(\d X_{1}^{*}-(\psi_{1}-\psi_{2})\psi^{*}_{1})
\exp{[X_{2}-X_{1}]}(z), 
\
\bet^{*}_{1}(z)=\psi^{*}_{1}\exp{[X_{2}-X_{1}]}(z)
\label{4.btgm01}
\enr
At the second substep we calculate $Q^{+}_{2}$-cohomology.

 On the equal footing one can take $Q^{+}_{2}+D_{orb}$-cohomology
as the first substep and apply $Q^{+}_{1}$ at the second substep.
Going by this way we obtain another $bc\bet\gm$ fields: 
\ber
\tld{b}_{0}(z)=\exp{[2X_{1}]}(z),
\
\tld{\bet}_{0}(z)=2\psi_{1}\exp{[2X_{1}]}(z),
\nmb
\tld{b}^{*}_{0}(z)=({1\ov 2}(\d X_{1}^{*}+\d X_{2}^{*})-
\psi_{1}(\psi^{*}_{1}+\psi^{*}_{2}))
\exp{[-2X_{1}]}(z), 
\nmb
\tld{\bet}^{*}_{0}(z)={1\ov 2}(\psi^{*}_{1}+\psi^{*}_{2})\exp{[-2X_{1}]}(z),
\nmb
\tld{b}_{1}(z)=\exp{[X_{2}-X_{1}]}(z),
\
\tld{\bet}_{1}(z)=(\psi_{2}-\psi_{1})\exp{[X_{2}-X_{1}]}(z),
\nmb
\tld{b}^{*}_{1}(z)=(\d X_{2}^{*}-(\psi_{2}-\psi_{1})\psi^{*}_{2})
\exp{[X_{1}-X_{2}]}(z), 
\nmb
\tld{\bet}^{*}_{1}(z)=\psi^{*}_{2}\exp{[X_{1}-X_{2}]}(z)
\label{4.btgm02}
\enr

 In view of the important property (\ref{3.coordtr1}) these two $bc\bet\gm$ systems are related to each other like the coordinates
of the standard covering of the total space of $O(2)$-bundle over $\mathbb{P}^{1}$
do
\ber
b_{0}=\tld{b}_{0}(\tld{b}_{1})^{2},
\
b_{1}=\tld{b}_{1}^{-1},...
\label{4.coordrel}
\enr
Therefore 
\ber
b_{0}(z)\leftrightarrow \ coordinate \ along \ the \ fiber \ b_{0},
\nmb
b_{1}(z) \leftrightarrow \ coordinate \ along \ the \ base \ b_{1}
\label{4.fieldcoord}
\enr
in the first open set of the standard covering. The tilda-fields 
service the second open set. Thus, the fields (\ref{4.btgm01}) and (\ref{4.btgm02})
generate the sections of the chiral de Rham complex over the open sets of the covering
given by the fan $\sgm_{1}\cup \sgm_{2}$.
Doing the second substep we calculate the cohomology of the Chech complex 
of the standard covering. It glues the sections of chiral de Rham complex over the open sets
into the chiral de Rham complex over the total space of the bundle. It is the end of the first
step of the cohomology calculation.

 The differential $Q^{-}$ of the second step cohomology calculation commutes with $D_{orb}$
and survives $\mathbb{Z}_{4}$-projection. It defines the function (potential) $W$ on the 
total space of $O(2)$-bundle and $Q^{-}$-cohomology calculation restricts the
chiral de Rham complex to the $dW=0$ point set of the function. In terms of the fields
(\ref{4.btgm01}) the potential takes the form
\ber
W=b_{0}^{2}(1+b_{1}^{4})
\label{4.W01}
\enr
The $dW=0$ points ($Q^{-}$-cohomology) are given by the equations
\ber
b_{0}=0, \ when \ b_{1}^{4}\neq -1,
\nmb
(b_{0})^{2}=0, \ when \ b_{1}^{4}=-1,
\label{4.dW01}
\enr
The set of solutions is $\mathbb{P}^{1}$ with 4 marked points $b_{1}^{4}=-1$,
where the additional states are possible according to the last row of (\ref{4.dW01}).
Thus, one can think of the $\mathbb{P}^{1}$ as a target
space of the model where the center of mass of the string is allowed to move.

 This interpretation is not quite correct however because we did not resolve the orbifold singularity completely. One can easy to see 
from (\ref{2.Vird}), (\ref{3.min}), (\ref{4.btgm01}) or (\ref{4.btgm02}) that the subgroup 
\ber
\mathbb{Z}_{2}\subset \mathbb{Z}_{4},
\label{4.fiberorb}
\enr
is acting on the sections of the chiral de Rham complex over the each open set.
But the action is nontrivial only along the fibers of the $O(2)$-bundle
so the base $\mathbb{P}^{1}$ is the fixed point set of the action. 
Therefore we should consider the target space of the model as 2 copies of $\mathbb{P}^{1}$
(except probably the points $b_{1}^{4}=-1$), where the second copy comes from the twisted sector.
This picture is in agreement with the result of \cite{FrSc} where the chiral 
de Rham complex on the orbifolds has been introduced. It was shown there that twisted sectors
chiral de Rham complex are the sheaves supported on the fixed points of the orbifold group
action. 

 Thus, the natural suggestion is that we reproduce the
geometry of 2-torus  which double covers the $\mathbb{P}^{1}$ with ramification
along the marked points $b_{1}^{4}=-1$. It is evidently confirmed by the Hodge numbers
calculation from (\ref{2.ZRR}): $h^{0,0}=h^{1,0}=h^{0,1}=h^{1,1}=1$.
Hence, adding the fermionic screening charge (\ref{4.Dorb})we blow up
the orbifold singularity of the Gepner model and obtain the $\sigma$-model on 2-torus which double
covers $\mathbb{P}^{1}$. 

\vskip 10pt
\leftline{\it 4.2. $K>2$ generalization.} In general case one has to deform $Q^{+}$ differential
(\ref{3.Qplus}) adding the screening charge
\ber
Q^{+}\rightarrow Q^{+}+D_{orb},
\nmb
D_{orb}=
\oint dz {1\ov K}(\psi^{*}_{1}+...+\psi^{*}_{K})\exp({1\ov K}(X^{*}_{1}+...+X^{*}_{K}))(z) 
\label{4.DorbK}
\enr
which comes from the spectral flow operator $\prod_{i=1}^{K}U_{i}$.
Similar to the $K=2$ case there are also another fermionic screening charges commuting
with $N=2$ Virasoro superalgebra currents as well as with the charges $Q^{-}_{i}$ 
but they do not appear as marginal operators of the model (\ref{2.ZRR}).

 The set of screening charges $\lbrc Q^{+}_{1},...,Q^{+}_{K},D_{orb}\rbrc$ defines the standard fan of 
the $O(K)$-bundle total space over $\mathbb{P}^{K-1}$.
The highest dimensional cones $\sgm_{i}$ of the fan are labeled by the differentials 
\ber
D_{i}=Q^{+}_{1}+...+Q^{+}_{i-1}+D_{orb}+Q^{+}_{i+1},...,Q^{+}_{K},
\ i=1,...,K
\label{4.Dcones}
\enr
where $Q^{+}_{i}$ is missing. 
In the standard basis $(e_{1},...,e_{K})$ of $\mathbb{R}^{K}$ the cones are generated by the
set of vectors $\Sgm_{i}$
\ber
\Sgm_{i}=(s_{1}=e_{1},...,s_{i-1}=e_{i-1},s_{i}={1\ov K}(e_{1}+...+e_{K}),
s_{i+1}=e_{i+1},...,s_{K}=e_{K})
\label{4.icones}
\enr
Making the first substep of the cohomology calculation we obtain a $bc\bet\gm$-system of
fields associated to each differential $D_{i}$ and the space of states generated by this
system is the set of sections of chiral de Rham complex over the open set associated to the cone $\sgm_{i}$ of the standard covering of the $O(K)$-bundle total space over $\mathbb{P}^{K-1}$. 
The analog of the formulas (\ref{4.btgm01}) can be written easily in terms of the 
dual basis $\check{\Sgm}_{i}$ to the $\Sgm_{i}$
\ber
\check {\Sgm}_{i}=(w_{(i)1},...w_{(i)K}),\
<w_{(i)j},s_{m}>=\dlt_{jm},
\label{4.dualcone}
\enr
Then the cohomology of $D_{i}$ is generated by
\ber
b_{(i)j}(z)=
\exp{[w_{(i)j}\cdot X]}(z),
\
\bet_{(i)j}(z)=w_{(i)j}\cdot\psi\exp{[w_{(i)j}\cdot X]}(z),
\nmb
b^{*}_{(i)j}(z)=(s_{j}\cdot \d X^{*}-
w_{(i)j}\cdot\psi s_{j}\cdot\psi^{*})
\exp{[-w_{(i)j}\cdot X]}(z), 
\nmb
\bet^{*}_{(i)j}(z)=s_{j}\cdot \psi^{*}\exp{[-w_{(i)j}\cdot X]}(z),
\label{4.btgmI}
\enr
where
\ber
b_{(i)i}(z)\leftrightarrow \ coordinate \ along \ the \ fiber, \ b_{(i)i}
\nmb
b_{(i)j}(z),\ j\neq i \leftrightarrow \ coordinates \ along \ the \ base \ b_{(i)j}
\label{4.fcoordK}
\enr

 The global sections of the chiral de Rham complex on the $O(K)$-bundle total
space are given by Chech complex associated to the standard covering \cite{B}.
It finishes the first step of cohomology calculation.

 In terms of the fields
(\ref{4.btgmI}) the LG potential determined by the differential $Q^{-}$ takes the form
\ber
W=(b_{(i)i})^{2}(1+\sum_{j\neq i}(b_{(i)j})^{2K})
\label{4.WI}
\enr
The $dW=0$ points ($Q^{-}$-cohomology) are given by the equations
\ber
b_{(i)i}=0, \ when \ \sum_{j\neq i}(b_{(i)j})^{2K}\neq -1,
\nmb
(b_{(i)i})^{2}=0, \ when \ \sum_{j\neq i}(b_{(i)j})^{2K}=-1,
\label{4.dWI}
\enr
Thus the set of solutions is $\mathbb{P}^{K-1}$ with marked submanifold 
\ber
\sum_{j\neq i}(b_{(i)j})^{2K}=-1,
\label{4.ramW}
\enr
where the additional states are possible according to the last row of (\ref{4.dWI}).

 Similar to the case $K=2$ one can see that only the fields of the fiber
are charged with respect to the operator $J[0]$ and the subgroup 
$\mathbb{Z}_{2}\subset \mathbb{Z}_{2K}$ is acting non-trivialy along the fibers.
Thus, the base $\mathbb{P}^{K-1}$ (considering as a zero section of the $O(K)$-bundle)
is the fixed point set of the $\mathbb{Z}_{2}$-action and we conclude that the 
target space of the model is 2 copies of $\mathbb{P}^{K-1}$
(except the submanifold (\ref{4.ramW})), where the second copy comes from the twisted
sector (see \cite{FrSc}). 

Hence, it is natural to suggest that the geometry of the model 
is the $K-1$-dimensional CY manifold geometry which 
double covers the $\mathbb{P}^{K-1}$ with ramification along the 
submanifold (\ref{4.ramW}). It is evidently confirmed by the Hodge numbers 
calculation from (\ref{2.ZRR}). For example, when $K=3$ 
\ber
h^{0,0}=h^{2,0}=h^{0,2}=h^{2,2}=1,\ h^{1,1}=20
\label{4.hK3}
\enr
which are Hodge numbers of $K3$. When $K=4$ we find
\ber
h^{0,0}=h^{3,0}=h^{0,3}=h^{3,3}=h^{1,1}=h^{2,2}=1,
\nmb
h^{1,2}=h^{2,1}=149
\label{4.hCY}
\enr
Thus, adding the fermionic screening charge (\ref{4.DorbK}) we blow up
the orbifold singularity of the Gepner model and obtain the $\sigma$-model on the CY manifold
which double covers $\mathbb{P}^{K-1}$. 

 It is important to note that in our free-field realization the center 
of mass of the string is allowed to move on the $\mathbb{P}^{K-1}$ 
which can be considered as the target space
and hence we can interprate the model as a $\sgm$-model on $\mathbb{P}^{K-1}$.
Though, the target space is not a CY manifold, nevertheless we have $N=2$ superconformal invariance. 
The possible solution of this puzzle is to consider these models as the examples of flux compactification 
\cite{DugK}, \cite{Gran}. Moreover, the models considered here are very close to
the known examples of the weak coupling limit of F-theory compactifications \cite{Sen1}, \cite{Sen2}.
The only difference is that they do not have the orientifolds planes.
It is interesting to know if these models can be related with F-theory compactifications.

 Finishing the Section we mention the question what is geometry of mirror
models. It can be investigated by free-field $bc\bet\gm$-representation but
we left it for the future.
 
\vskip 20pt
\leftline{\bf Acknowledgments}
\vskip 10pt

I thank Lev Borisov and Feodor Malikov for discussions.
This work was conducted in part within the framework of the federal 
program "Scientific and Scientific-Pedagogical Personnel of Innovational Russia" (2009-2013), by the RFBR initiative interdisciplinary project (Grant No. 09-02-
12446-ofi-m) as well as by grants RFBR-07-02-00799-a, SS3472.2008.2,
RFBR-CNRS 09-02-93106

\end{document}